# Artificial Intelligence as Game Changer in Cybersecurity: What We Learned in 2025-2026, and how this is relevant for Africa

*The Decisive Instrument, the African Exclusion, and the Choices That Remain*

Mikael Alemu Gorsky

Holon Institute of Technology

Abstract

In 2025 and 2026, two events settled questions that had until then been speculative. In the first, a large language model executed the great majority of a state-aligned cyber-espionage campaign autonomously, with human operators intervening only at a few strategic decision points; the model discovered vulnerabilities, exploited them, moved laterally, and exfiltrated data largely on its own. In the second, quieter event, the most capable cyber-relevant model was placed under a controlled-access program limited to a vetted set of United States technology firms, allied governments, and European standards bodies. That perimeter includes no African government, operator, or university. Read together, the two events establish the argument of this paper: frontier language models have become a decisive instrument of cyber operations, and that instrument is built, owned, and rationed within a small circle from which Africa is absent.

The paper shows that Africa stands outside the perimeter on every count. It does not build frontier models; it cannot yet operate them; and it cannot, for now, obtain the most capable ones. The operational deficit is documented along three axes — skilled people, compute and electrical power, and investment — each set against current figures, including the latest national workforce counts, the continent's share of global data-center capacity, and quarterly AI-investment flows. The same instrument is already turning against the continent: AI-enabled fraud is mounting against African mobile-money systems, the part of the digital economy the continent leads, and imported defensive tools fit local conditions poorly.

Two distinct constraints are then examined. The first is the gating of frontier models by their developers, a constraint no African decision can open. The second is the dependence of African networks on infrastructure vendors now caught in geopolitical restriction. African states chose this dependence deliberately, on rational terms at the time, and they can still revisit it. Because the developers themselves forecast that comparable but ungated models will proliferate within six to twelve months, far faster than any domestic capacity can be built, the paper argues that the only realistic response operates inside that short window: entering threat-intelligence sharing arrangements, adopting governance frameworks already authored elsewhere, and opening partnership negotiations with those who hold the instrument. The continent enters such collaboration as a contributor of data, conditions, and legitimacy, not as a petitioner. What it makes of the window will be decided by Africans.

## Introduction: the instrument arrives, and the gate closes

In mid-September 2025, the AI company Anthropic detected a cyber espionage operation that it later assessed, with high confidence, to be the work of a Chinese state-sponsored group it designated GTG-1002 [1]. The operation was unremarkable in its targets, roughly thirty technology firms, financial institutions, chemical manufacturers, and government agencies, and a handful of confirmed intrusions [1]. What made it a turning point was who did the work. The attackers manipulated Claude Code, an agentic coding tool, into believing it was performing authorized defensive testing, and then let it run. The model executed an estimated eighty to ninety percent of the tactical operations on its own, discovering vulnerabilities, exploiting them, moving laterally, escalating privilege, and exfiltrating data, while the human operators intervened only at a few strategic decision points [1]. Anthropic's own conclusion was blunt: the barriers to performing sophisticated cyberattacks have dropped substantially, and a threat actor with the right setup can now use an agentic system to do the work of an entire team of experienced hackers [1]. The United States House Committee on Homeland Security found the episode serious enough to summon Anthropic's chief executive to testify [2].

GTG-1002 settled a question that had been theoretical until then. The large language model is no longer an assistant to the human attacker; it is the attacker, supervised. The instrument has arrived on the field of cyber operations, and it works.

A second event, two months earlier and quieter, settled a different question. In the autumn of 2025 Anthropic placed its most advanced and unreleased model, designated Mythos, into the hands of a small group of large technology and cybersecurity firms, because it judged that the model could raise the likelihood of large-scale AI-driven cyberattacks and wanted its defensive value in trusted hands before that danger matured [3]. Access was conditioned on meeting security requirements, and the early membership describes the perimeter precisely: Amazon Web Services, Apple, Broadcom, Cisco, CrowdStrike, Google, JPMorgan Chase, Microsoft, and Nvidia, later joined by NATO and the European Union's cybersecurity agency, ENISA [3]. No African government, operator, or university appears in that membership. Anthropic added the forecast that gives the whole arrangement its urgency: within six to twelve months, it expects, many other companies will hold models of comparable capability, and some may release them without the safeguards that prevent misuse [3].

The first event shows what the instrument can do. The second shows who controls it and on what terms. Read together, they state the argument of this paper. In 2025 and 2026 large language models became the decisive instrument of cyber operations, that instrument is held and rationed by a small number of actors in a single country, and Africa sits outside the perimeter in every capacity at once: it cannot build the instrument, it cannot yet wield the instrument, and it cannot, for now, obtain the instrument. We must accept the reality: today Africa is neither the developer of frontier LLMs nor a member of closed circuit of pioneer users of LLMs.

The sections that follow establish each part of that claim and then ask what an African response can realistically be, given that the most dangerous version of the instrument is forecast to spread within the year.

## 1. This instrument is decisive, and the most advanced tools are made in USA

Two propositions define the strategic position. The first is that artificial intelligence has become a genuine weapon in cyber operations, not an aid to one. The second is that the frontier of this weapon belongs to U.S. laboratories.

Consider first what the leading models can now do against human experts in the hardest formal contests, because cyber capability is one instance of a more general reasoning capability, and the general case is easier to measure cleanly. At the 2025 International Collegiate Programming Contest World Finals, the most demanding programming competition in the world, an OpenAI system solved all twelve problems and a Google DeepMind system solved ten, under the same five-hour limit and the same real-time judging applied to the human teams, with no contest-specific tuning [4]. A perfect score at the world championship of human programming is not an incremental gain over the previous year; it is the disappearance of a ceiling. The same general systems reached gold-medal standard at the International Mathematical Olympiad and the International Olympiad in Informatics within the same season, the informatics result placing ahead of all but a handful of the world's strongest teenage competitors [5]. Reasoning at the level of the best humans in mathematics and programming is no longer a forecast.

Pointed at security specifically, the same capability shows the same curve. The United Kingdom's AI Security Institute reports that two years ago the best available models could barely complete beginner-level cyber tasks, while on expert-level capture-the-flag challenges that no model could finish before April 2025, the Mythos preview now succeeds roughly seventy-three percent of the time [3]. On CyberGym, the largest benchmark of real-world vulnerability discovery, the same model scored above eighty percent against a previous published best near twenty percent, an improvement of nearly four times in a single generation [6]. A tool that finds and exploits real software vulnerabilities four times more often than the prior state of the art, in the space of one release cycle, is the technical fact beneath GTG-1002.

The second proposition, that the most advanced LLMs are made in USA, rests on the frontier of capability rather than on the broad middle of the market, where Chinese open-weight models have largely closed the gap. On the four hardest reasoning benchmarks, the leaders are all United States laboratories. On FrontierMath, a benchmark of unpublished problems built by professional mathematicians including Fields Medalists, OpenAI's GPT-5.5 Pro now solves about 52% of the set, where a year earlier the best model managed roughly 25% [7]. On Humanity's Last Exam, billed as the hardest general benchmark available, Anthropic's Claude Mythos Preview now answers about 65% of the questions correctly, where the best models scored under 3% when the test was released in early 2025 [7]. On GPQA Diamond, which tests doctoral-level scientific reasoning, Anthropic's Claude Mythos Preview and Google's Gemini 3.1 Pro both score about 94%, well above the roughly 65% that PhD experts reach in their own fields [8]. On ARC-AGI-2, a test of novel abstract reasoning on which every frontier model scored 0% at its launch in March 2025, OpenAI's GPT-5.5 now reaches about 85%, above the 66% that the average person scores [9]. The Chinese models that match Western performance do so on well-defined coding tasks and at far lower cost, not at the reasoning frontier, and they share one decisive property: they are open-weight, downloadable and self-hostable by anyone. The United States frontier models that lead these benchmarks, and in particular the cyber-capable models gated through programs like the one around Mythos, are closed. The part of the frontier that can be controlled is American and shut; the part that cannot be controlled is open and flows to everyone, including every attacker.

The concentration is visible in money as well as in scores. In 2025, AI-related venture investment roughly doubled to about $259 billion, with three-quarters of it flowing to United States companies [10]. United States private investment in artificial intelligence reached $285.9 billion in the same year, against China's $12.4 billion, a gap of more than twenty to one [8]. The honest qualification is that private figures understate China's true effort, which runs heavily through state guidance funds the private numbers do not capture, and that the quality gap between the best American and Chinese models has narrowed to a few percentage points [8]. The narrowing matters, and it sharpens rather than softens the argument: the present American lead at the closed frontier is a window, not a permanent condition. For the duration of that window, which is the period this paper concerns, the decisive instrument is built, owned, and rationed in the United States.

## 2. Africa cannot wield the instrument, and is already its target

A weapon one cannot build might still be wielded if one had the means to operate it. Africa does not yet have those means, and the deficit can be measured along three axes: the people, the physical substrate of compute and power, and the capital. Each is stark on its own, and together they explain why the continent cannot yet field the instrument within the window that matters. The gap is not one of will or talent; it is one of people, power, and capital, the three things the instrument demands.

### *Human capital*

The workforce gap is the most familiar of the three and still the most severe. In the most recent national counts available, ISC2 put Nigeria at 8,352 cybersecurity professionals and South Africa at 57,269, against a United States workforce of 482,985 [11]. Those figures date to 2023 and remain the latest country-level estimates ISC2 has published, the firm having since dropped per-country headcounts; on the continent's growth trajectory they understate today's numbers, yet even generous revision leaves the gap vast. Across the continent, fewer than 300,000 professionals are available to defend one of the world's fastest-growing digital economies, and the shortfall runs to tens of thousands of unfilled roles [12]. The pipeline behind those numbers is the deeper problem: only about eleven percent of tertiary graduates on the continent have received any formal digital training [12]. A defensive AI tool, however capable, does not operate itself; it requires skilled people to deploy, tune, and supervise it, and skilled people remain in critically short supply. The instrument that lets one operator do the work of a team is most valuable exactly where teams are hardest to staff, and least accessible there for the same reason. The professionals already carrying that load feel the symmetry most sharply.

### *Compute and energy*

The instrument runs on computation, and computation runs on power. Africa is short of both. The continent holds under one percent of global data-center capacity while housing eighteen percent of the world's population [13]. That single ratio fixes the physical position: the machines that run frontier models are overwhelmingly located elsewhere, and the small share that exists on the continent is concentrated in a few coastal markets.

The shortage of compute cannot be separated from the shortage of electricity, because the second binds the first. Almost 600 million people in Africa lack access to electricity today [14]. Sub-Saharan Africa now accounts for the large majority of the global population without power [15]. The scale of the generation deficit becomes concrete when the continent's grids are set against the machines they would have to power: Ethiopia's entire national generation capacity stands at roughly 9,500 megawatts, while a single new U.S. AI data-center such as Vantage's planned Frontier site in Texas is designed for 1,400 megawatts, and OpenAI's Stargate initiative is committed to 10,000 megawatts in all — more than Ethiopia's whole grid. Where capacity exists on paper, it often fails in practice; in Nigeria only a quarter of installed generation is actually available because of constraints in generation, transmission, and distribution [16]. Against this background, the demand that frontier AI places on the grid is severe and growing, because the processors that run these models are unusually power-hungry and global data-center consumption is set to double by 2030 [17].

### *Investment*

The capital gap is the sharpest of the three, and it connects directly to the concentration described in the previous section. United States AI companies raised about $159 billion in 2025, roughly seventy-nine percent of global AI startup funding [18]. The entire African total, by contrast, is a rounding error against that figure. One quarterly snapshot makes the disparity concrete: in the second quarter of 2025, African AI startups raised about $14 million across five deals, equal to roughly two-hundredths of one percent of the $47.3 billion invested globally in the same period, a quarter in which United States companies alone took nearly $40 billion [19]. Measured over five years rather than one quarter, African AI startups had raised, by mid-2025, a cumulative total smaller than what a single leading United States laboratory raises in one funding round [20].

The same concentration that funds the frontier is actively draining capital away from the continent. As the global AI boom pulls venture money toward the United States, African founders have been forced to turn from international venture capital toward domestic development-finance institutions, pension funds, and debt [10]. So the same boom that builds the instrument is draining the capital Africa would need to respond to it, and the gap on all three axes is widening just as the instrument that exploits it arrives.

That arrival is not hypothetical. The instrument is already being turned on Africa, and the evidence is in the part of the digital economy the continent leads. Africa's mobile-money system, worth on the order of $1.4 trillion in annual transaction value, is under direct pressure from AI-enabled fraud, with attackers using generated identities, forged documents, and synthetic biometric artifacts to defeat customer-verification checks at scale [21]. The figures move in one direction: deepfake-related incidents in South Africa rose more than 260 percent year on year, and in Kenya such attacks now account for nearly a tenth of all fraud attempts, even as fraud rates fell in the United States and the European Union and rose in Africa [22, 23]. The structural twist is that imported defensive tools, trained on other markets' data and other languages' patterns, fit African conditions poorly, so even purchased capability underperforms locally [21]. The instrument lands hardest on the markets least equipped, so far, to answer it.

## 3. The exclusion is structural, and is supported by two very different locks

Africa cannot build the instrument and cannot yet wield it. The remaining question is whether it can obtain the instrument from those who hold it. The answer involves two locks of very different character.

### *The 'gating' lock*

The program around Mythos, described in the introduction, distributes the most capable cyber-relevant model to a vetted set of institutions that meet the gatekeeper's security requirements, and that set is drawn along a clear geopolitical line: United States technology firms, allied governments, and European standards bodies [3]. No African entity is inside it. This exclusion is not the product of an African failure to apply or negotiate; it is a sovereign decision made by a private United States company under United States strategic conditions, and no policy adopted in Addis Ababa, Nairobi, or Pretoria can alter it. Africa's position relative to this lock is simply to be outside it.

The forecast attached to the program makes the lock's timing the central danger. The gate holds the instrument inside a trusted perimeter only for as long as the capability remains rare. Once equivalent models proliferate, which their builders expect within six to twelve months, the instrument reaches everyone, and the first to benefit are not under-resourced defenders but attackers and ungated actors who face no approval process at all [3]. The window during which the instrument is both decisive and controlled is therefore also the window during which Africa is locked out of the controlled version while remaining fully exposed to the uncontrolled one.

### *The 'vendor' lock*

The second lock is the dependency of African digital infrastructure on Chinese vendors. Chinese firms built much of the continent's connective tissue. Huawei alone constructed an estimated seventy percent of Africa's 4G networks and leads 5G expansion across roughly thirty markets, embedding equipment deep in critical national infrastructure, while ZTE holds particular weight in Ethiopia and elsewhere, and many e-government platforms and national-security communications depend on Chinese systems to function [24, 25]. In Ethiopia specifically, a Huawei partnership with the state operator raised 4G capacity by seventy percent [23]. This dependency now sits inside a vise: the same Western bloc that gates the frontier model also restricts these vendors, as the United States 2025 Annual Threat Assessment named the capacity of Chinese actors to disrupt foreign telecommunications infrastructure as a coercive instrument [25]. The infrastructure that most needs the defensive instrument is operated on equipment supplied by firms barred from that instrument.

African states chose these vendors, deliberately and repeatedly, and in several cases continue to choose them on explicit political grounds. The choice was rational on the terms that applied when it was made across the 2010s: Chinese suppliers offered competitive equipment, faster build-out, and financing that Western competitors would not match, against a real shortage of local champions and capital [24]. Nothing in that calculation was foolish. The single variable that now makes the choice costly, the arrival of a gated, geopolitically aligned frontier instrument that turns the supplier's national alignment into a defensive liability, did not exist when the contracts were signed. Both halves of the relevant condition, the decisive instrument and its gating, are phenomena of 2025 and 2026.

## 4. Conclusion: an operational response inside the window

The events of 2025 and 2026 belong together. The arrival of a gated, frontier-grade cyber instrument created a new class of threat, left Africa outside the protected perimeter, and rewrote the value of a decade of infrastructure procurement — all within roughly eighteen months, and all from the same cause.

The danger this poses to Africa is not chronic but immediate, and the distinction governs everything that follows. The builders of the instrument forecast that ungated, comparably capable models will proliferate within six to twelve months [3]. The capacity deficits described above, in people, in compute and power, and in capital, are the work of years to close. Two clocks are running at once, and they are set to different scales: the threat advances on a calendar of months, the capacity to answer it on a calendar of years. The exposure is the gap between them. Any response pitched to the slower clock fails by construction.

This rules out an entire class of well-meaning answers. A ten-year plan for AI education, a sovereign-compute build-out, a homegrown frontier laboratory, each may be worth pursuing for its own sake, and none of them addresses a threat that arrives before any of them matures. To offer such plans as the response to this danger is not strategy but evasion, because it answers a slower question than the one actually posed. The constraint is severe and it is clarifying: the response must be operational within 6-12 months, the same window the instrument's own makers have named, or it is not a response to this problem at all.

Only one class of action fits that window, and it is not the building of new capability but the borrowing of capability that already exists, through collaboration with those who hold it.

Nothing built from scratch matures in months; agreements to share and integrate what is already built can. The realistic, near-term core of an African response is therefore threefold:

First, the continent can enter threat-intelligence sharing arrangements with international organizations already operating at the required scale, an extension of the model already proven by coordinated continental enforcement, which has dismantled thousands of fraud operations on timelines of months rather than years [26].
Second, it can adopt the safety and governance frameworks that European institutions, themselves inside the trusted perimeter, have already authored, rather than spending years drafting its own.
Third, it can open partnership negotiations with the laboratories and corporations that hold the frontier instrument, the harder and slower of the three tracks, but one that the first two position the continent to pursue. The actors named, international organizations, global non-governmental bodies, and the leading firms, are precisely those who can move at the speed the window demands, because for them the task is connection, not creation.
The argument of this paper has been hard, and its conclusion is not. Africa cannot out-build the threat in the time it has, so it must act now, through the partnerships already within reach, and it must do so without apology. A continent of more than a billion people, on the front line of the fraud these systems enable and holding data no laboratory can gather from the outside, comes to the table with something to offer, not with empty hands. The window is narrow and it is closing. What Africa makes of it will be decided by Africans, and by what it will bring to the table while there is still a table to join.